\documentclass{article}
\usepackage[preprint]{spconfa4}
\usepackage{amsmath,graphicx,siunitx,bbm}
\usepackage{pgfplots, tikz} 
\usepgfplotslibrary{groupplots}
\pgfplotsset{compat=1.12}
\usepackage{csvsimple} 
\usepackage{hyperref} 
\usepackage{float}
\usepackage{booktabs, makecell, multirow} 
\usepackage{bigdelim} 
\usepackage{enumitem} 

\def\etal{et al.~}

\newcommand{\Fig}[1] {Fig.~\ref{#1}}
\newcommand{\Tab}[1] {Table~\ref{#1}}

\newcommand{\nr}[0] {\multicolumn{1}{c}{-}} 
\newcommand{\rot}[1] {\rotatebox[origin=c]{90}{#1}}

\definecolor{bblue}{HTML}{4F81BD}
\definecolor{rred}{HTML}{C0504D}
\definecolor{ggreen}{HTML}{9BBB59}
\definecolor{ppurple}{HTML}{9F4C7C}
\pgfkeys{/pgf/number format/.cd,1000 sep={}}

\usepackage{etoolbox}
\patchcmd{\thebibliography}
  {\settowidth}
  {\setlength{\itemsep}{0pt plus 0.5pt}\settowidth}
  {}{}

\copyrightnotice{978-1-6654-6867-1/22/\$31.00~\copyright2022 European Union}
                   
\title{DeepFilterNet2: Towards Real-Time Speech Enhancement on Embedded Devices for Full-Band Audio}
\twoauthors
{H. Schröter, A. Maier\sthanks{A. Maier is the last author of this paper.}}
{Pattern Recognition Lab\\
Friedrich-Alexander-Universit\"at Erlangen-N\"urnberg\\
Erlangen, Germany}
{A.N. Escalante-B., T. Rosenkranz}
{WS Audiology\\
Research and Development\\
Erlangen, Germany}
\begin{document}
%
\maketitle

\begin{abstract}
Deep learning-based speech enhancement has seen huge improvements and recently also expanded to full band audio (\SI{48}{\kHz}).
However, many approaches have a rather high computational complexity and require big temporal buffers for real time usage e.g.~due to temporal convolutions or attention.
Both make those approaches not feasible on embedded devices.
This work further extends DeepFilterNet, which exploits harmonic structure of speech allowing for efficient speech enhancement (SE).
Several optimizations in the training procedure, data augmentation, and network structure result in state-of-the-art SE performance while reducing the real-time factor to \num{0.04} on a notebook Core-i5 CPU.
This makes the algorithm applicable to run on embedded devices in real-time.
The DeepFilterNet framework can be obtained under an open source license.
\end{abstract}

\section{Introduction}
\label{sec:intro}

Recently, deep learning-based speech enhancement have been extended to full-band (\SI{48}{\kHz}) \cite{valin2020perceptually,schroter2022deepfilternet,zhao2022frcrn,yu2022dmf}.
Most SOTA methods perform SE in frequency domain by applying a short-time Fourier transform (STFT) to the noisy audio signal and enhance the signal in an U-Net like deep neural network (DNN).
However, many approaches have relatively large computational demands in terms of multiply-accumulate operations (MACs) and memory bandwidth.
That is, the higher sampling rate usually requires large FFT windows resulting in a high number of frequency bins which directly translates to a higher number of MACs.

PercepNet \cite{valin2020perceptually} tackles this problem by using a triangular ERB (equivalent rectangular bandwidth) filter bank. Here, the frequency bins of the magnitude spectrogram are logarithmically compressed to \num{32} ERB bands.
However, this only allows real-valued processing which is why PercepNet additionally applies a comb-filter for finer enhancement of periodic component of speech.
FRCRN \cite{zhao2022frcrn} instead splits the frequency bins into 3 channels to reduce the size of the frequency axis.
This approaches allows complex processing and prediction of a complex ratio mask (CRM).
Similarly, DMF-Net \cite{yu2022dmf} uses a multi-band approach, where the frequency axis is split into 3 bands that are separately processed by different networks.
Generally, multi-stage networks like DMF-Net have recently demonstrated their potential compared to single stage approaches.
GaGNet \cite{li2022glance}, for instance, uses two so called glance and gaze stages after a feature extraction stage.
The glance module works on a coarse magnitude domain, while the gaze module processes the spectrum in complex domain allowing to reconstruct the spectrum at a finer resolution.

In this work we extend the work from \cite{schroter2022deepfilternet} which also operates in two stages.
DeepFilterNet takes advantage of the speech model consisting of a periodic and a stochastic component.
The first stage operates in ERB domain, only enhancing the speech envelope, while the second stage uses deep filtering \cite{schroeter2020clcnet, mack2019deep} to enhance the periodic component.
In this paper, we describe several optimizations resulting in SOTA performance on the Voicebank+Demand \cite{valentini2016investigating} and deep noise suppression (DNS) 4 blind test challenge dataset \cite{dubey2022dns4}.
Moreover, these optimizations lead to an increased run-time performance, making it possible to run the model in real-time on a Raspberry Pi 4.

\section{Methods}
\label{sec:methods}

\begin{figure*}
  \begin{center}
    \includegraphics[width=\linewidth,trim=0 11cm 0 0.5cm, clip]{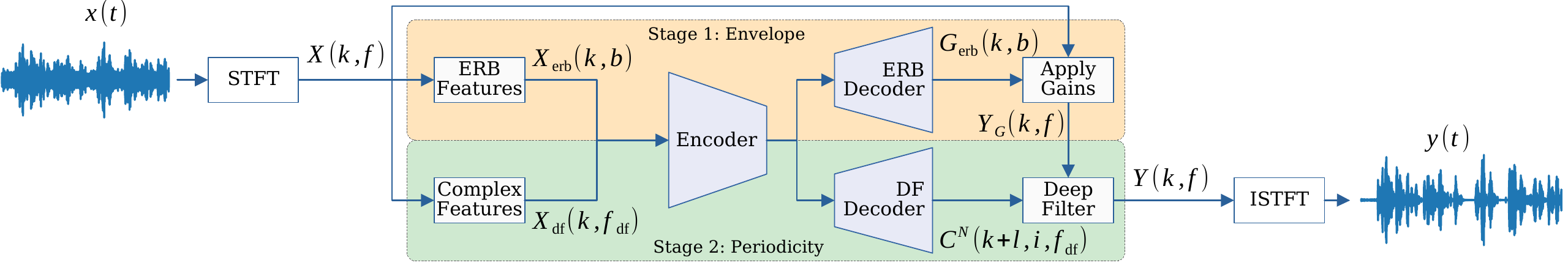}
  \end{center}
  \vspace{-1.5em}
  \caption{Schematic overview of the DeepFilterNet2 speech enhancement process.}
  \vspace{-1.0em}
  \label{fig:framework}
\end{figure*}

\subsection{Signal Model and the DeepFilterNet framework}
We assume noise and speech to be uncorrelated such as:
\vspace{-.25em}\begin{equation}
  x(t) = s(t) * h(t) + n(t)
\end{equation}
where $s(t)$ is a clean speech signal, $n(t)$ is an additive noise, and $h(t)$ a room impulse response modeling the reverberant environment resulting in a noisy mixture $x(t)$.
This directly translates to frequency domain:
\begin{equation}
  X(k, f) = S(k, f) \cdot H(k, f) + N(k, f)\text{,}
\end{equation}
where $X(k, f)$ is the STFT representation of the time domain signal $x(t)$ and $k$, $f$ are the time and frequency indices.

In this work, we adopt the two-stage denoising process of DeepFilterNet \cite{schroter2022deepfilternet}.
That is, the first stage operates in magnitude domain and predicts real-valued gains.
The whole first stage operates in an compressed ERB domain which serves the purpose of reducing computational complexity while modeling auditory perception of the human ear.
Thus, the aim of the first stage is to enhance the speech envelope given its coarse frequency resolution.
The second stage operates in complex domain utilizing deep filtering \cite{mack2019deep, schroeter2020clcnet} and is trying to reconstruct the periodicity of speech.
\cite{schroter2022deepfilternet} showed, that deep filtering (DF) generally outperforms traditional complex ratio masks (CRMs) especially in very noisy conditions.

The combined SE procedure can be formulated as follows. An encoder $\mathcal F_\text{enc}$ encodes both ERB and complex features into one embedding $\mathcal E$.
\begin{equation}
  \mathcal E(k) = \mathcal F_\text{enc}(X_\text{erb}(k, b), X_\text{df}(k, f_\text{erb})) \\
  \label{eq:enc}
\end{equation}
Next, the first stage predicts real-valued gains $G$ and enhances the speech envelope resulting in the short-time spectrum $Y_G$.
\begin{equation}
\begin{split}
  G_\text{erb}(k, b) &= \mathcal F_\text{erb\_dec}(\mathcal E(k)) \\
  G(k, f) &= \text{interp}(G_\text{erb}(k, b)) \\
  Y_G(k, f) &= X(k, f) \cdot G(k, f)
\end{split}
\end{equation}
Finally in the second stage, $F_\text{df\_dec}$ predicts DF coefficients $C_\text{df}^N$ of order $N$ which are then linearly applied to $Y_G$.
\begin{equation}
\begin{split}
  C^N_\text{df}(k, i, f_\text{df}) &= \mathcal F_\text{df\_dec}(\mathcal E(k)) \\
  Y(k, f') &= \sum_{i=0}^{N} C(k, i, f') \cdot X(k - i + l, f) \text{,}
\end{split}
\end{equation}
where $l$ is the DF look-ahead.
As stated before, the second stage only operates on the lower part of the spectrogram up to a frequency $f_\text{df}=\SI{5}{\kHz}$.
The DeepFilterNet2 framework is visualized in \Fig{fig:framework}.

\subsection{Training Procedure}

\begin{figure}
  \begin{center}
    \includegraphics[width=\linewidth]{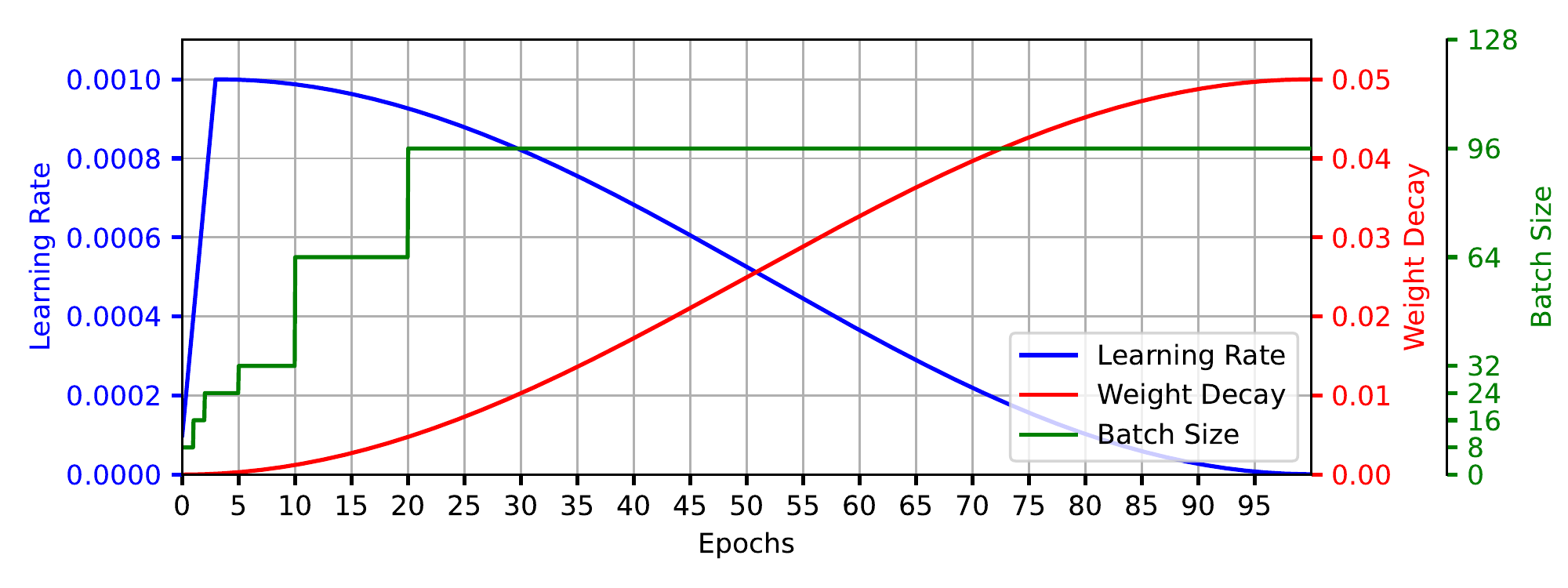}
  \end{center}
  \vspace{-2.0em}
  \caption{Learning rate, weight decay and batch size scheduling used for training.}
  \vspace{-0.5em}
  \label{fig:lrs}
\end{figure}

In DeepFilterNet \cite{schroter2022deepfilternet}, we used an exponential learning rate schedule and fixed weight decay.
In this work, we additionally use a learning rate warmup of 3 epochs followed by a cosine decay. Most importantly, we update the learning rate at every iteration, instead of after each epoch.
Similarly, we schedule the weight decay with an increasing cosine schedule resulting in a larger regularization for the later stages of the training.
Finally, to achieve faster convergence especially in the beginning of the training, we use batch scheduling \cite{smith2017don} starting with a batch size of 8 and gradually increasing it to 96.
The scheduling scheme can be observed in \Fig{fig:lrs}.

\subsection{Multi-Target Loss}

We adopt the spectrogram loss $\mathcal{L}_\text{spec}$ from \cite{schroter2022deepfilternet}. Additionally use a multi-resolution (MR) spectrogram loss where the enhancement spectrogram $Y(k, f)$ is first transformed into time-domain before computing multiple STFTs with windows from \SIrange{5}{40}{\ms} \cite{choi2021real}.
To propagate the gradient for this loss, we use the pytorch STFT/ISTFT, which is numerically sufficiently close to the original DeepFilterNet processing loop implemented in Rust.
\begin{equation}
  \mathcal{L}_\text{MR} =
  \sum_{i} ||\ |Y'_i|^c - |S'_i|^c ||^2 ||\ |Y'_i|^c e^{j\varphi_Y} - |S'_i|^c e^{j\varphi_S} ||^2\text{,}
\end{equation}
where $Y'_i = \text{STFT}_i(y)$ is the i-th STFT with window sizes in $\{5, 10, 20, 40\}\si{\ms}$ of the predicted TD signal $y$, and $c=0.3$ is a compression parameter \cite{valin2020perceptually}.
Compared to DeepFilterNet \cite{schroter2022deepfilternet}, we drop the $\alpha$ loss term since the employed heuristic is only a poor approximation of the local speeech periodicity.
Also, DF may enhance speech in non-voiced sections and can disable its effect by setting the real part of the coefficient at $t_0$ to 1 and the remaining coefficients to \num{0}.
The combined multi-target loss is given by:
\begin{equation}
  \mathcal{L} = \lambda_\text{spec} \mathcal{L}_\text{spec} + \lambda_\text{MR} \mathcal{L}_\text{MR}
  \label{eq:mtloss}
\end{equation}

\subsection{Data and Augmentation}
\label{ssec:data}

While DeepFilterNet was trained on the deep noise suppression (DNS) 3 challenge dataset \cite{reddy2021dns3}, we train DeepFilterNet2 on the english part of DNS4 \cite{dubey2022dns4} which contains more full-band noise and speech samples.\\
In speech enhancement, usually only background noise and in some cases reverberation is reduced \cite{valin2020perceptually, choi2021real, schroter2022deepfilternet}.
In this work, we further extended the SE concept to declipping. Therefore, we distinguish between \textit{augmentations} and \textit{distortions} in the on-the-fly data pre-processing pipeline.
Augmentations are applied to speech and noise samples with the aim of further extending the data distributions the network observes during training.
Distortions, on the other hand, are only applied to speech samples for noisy mixture creation. The clean speech target is not affected by a distortion transform. Thus, the DNN learns to reconstruct the original, undistorted speech signal.
Currently, the DeepFilterNet framework supports the following randomized \textit{augmentations}:
\begin{itemize}[noitemsep,topsep=0pt,parsep=0pt,partopsep=0pt]
  \item Random 2nd order filtering \cite{valin2018rnnoise} 
  \item Gain changes
  \item Equalizer via 2nd order filters
  \item Resampling for speed and pitch changes \cite{valin2018rnnoise}
  \item Addition of colored noise (not used for speech samples)
\end{itemize}
Additionally to denoising, DeepFilterNet will try to revert the following \textit{distortions}:
\begin{itemize}[noitemsep,topsep=0pt,parsep=0pt,partopsep=0pt]
  \item Reverberation; the target signal will contain a smaller amount of reverberation by decaying the room transfer function.
  \item Clipping artifacts with SNRs in $[20, 0]\si{\dB}$.
\end{itemize}

\subsection{DNN}
\label{ssec:dnn}

\begin{figure}[tb]
  \begin{center}
    \includegraphics[width=\linewidth, page=2, trim=0 4cm 5.5cm 0, clip]{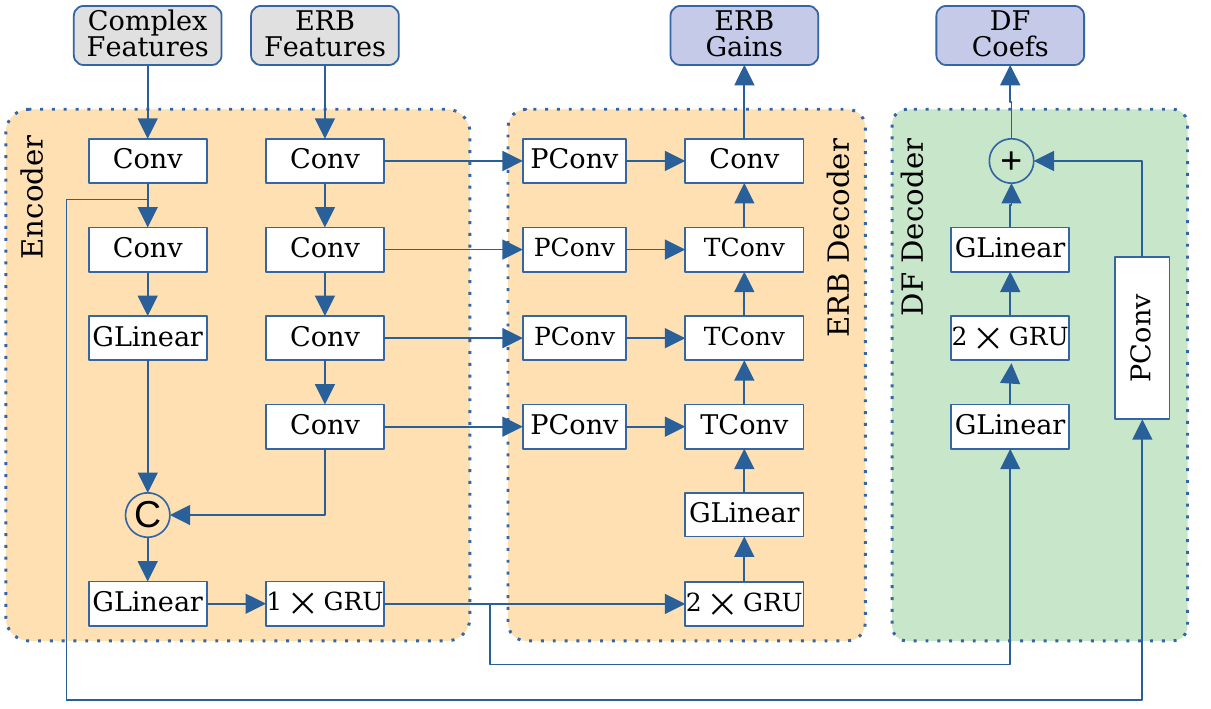}
  \end{center}
  \vspace{-1.5em}
  \caption{DeepFilterNet2 architecture.}
  \vspace{-1.2em}
  \label{fig:dnn}
\end{figure}

We keep the general convolutional U-Net structure of DeepFilterNet \cite{schroter2022deepfilternet}, but make the following adjustments.
The final architecture is shown in \Fig{fig:dnn}.
\begin{enumerate}[leftmargin=*,topsep=2pt,parsep=0pt,itemsep=4pt]
  \item \textit{Unification of the encoder}. Convolutions for both ERB and complex features are now processed within the encoder, concatenated, and passed to a grouped linear (GLinear) layer and single GRU.
  \item \textit{Simplify Grouping}. Previously, grouping of linear and GRU layers was implemented via separate smaller layers which results in a relatively high processing overhead. In DeepFilterNet2, only linear layers are grouped over the frequency axis, implemented via a single matrix multiplication. The GRU hidden dim was instead reduced to \num{256}. We also apply grouping in the output layer of the DF decoder with the incentive that the neighboring frequencies are sufficient for predicting the filter coefficients. This greatly reduces run-time, while only minimaly increasing the number of FLOPs.
  \item \textit{Reduction of temporal kernels}. While temporal convolutions (TCN) or temporal attention have been successfully applied to SE, they require temporal buffers during real-time inference. This can be efficiently implemented via ring buffers, however, the buffers need to be held in memory. This additional memory access may result in bandwidth being the limiting bottleneck, which could be the case especially for embedded devices. Therefore, we reduce the kernel size of the convolutions and transposed convolutions from $2\times3$ to $1\times3$, that is 1D over frequency axis. Only the input layer now incorporates temporal context via a causal $3\times3$ convolution. This drastically reduces the use of temporal buffers during real-time inference.
  \item \textit{Depthwise pathway convolutions}. When using separable convolutions, the vast amount of parameters and FLOPs is located at the $1\times1$ convolutions. Thus, adding grouping to pathway convolutions (PConv) results in a great parameter reduction while not losing any significant SE performance.
\end{enumerate}

\vspace{-0.5em}
\subsection{Post-Filter}

We adopt the post-filter, first proposed by Valin \etal\cite{valin2020perceptually}, with the aim of slightly over-attenuating noisy TF bins while adding some gain back to less noisy bins.
We perform this on the predicted gains in the first stage:
\begin{equation}
  \begin{split}
    G'(k, b) &\leftarrow G(k, b))\cdot\sin\Big(\dfrac{\pi}{2}G(k, b)\Big) \\
    G(k, b)  &\leftarrow \dfrac{(1+\beta) \cdot G(k, b)}{1+\beta+G'(b, k)} \text{\ .}
  \label{eq:postfilter}
  \end{split}
  \vspace{-1em}
\end{equation}

\section{Experiments}
\label{sec:experiments}
\subsection{Implementation details}
As stated in section \ref{ssec:data}, we train DeepFilterNet2 on DNS4 dataset using overall more than \SI{500}{\hour} of full-band clean speech, approx.~\SI{150}{\hour} of noise as well as \num{150} real and \num{60000} simulated HRTFs.
We split the data into train, validation and test sets (\SI{70}{\percent}, \SI{15}{\percent}, \SI{15}{\percent}).
The Voicebank set was split speaker-exclusive with no overlap with test set. We evaluate our approach on the Voicebank+Demand test set \cite{valentini2016investigating} as well as the DNS4 blind test set \cite{dubey2022dns4}.
We train the model with AdamW for \num{100} epochs and select the best model based on the validation loss.

In this work, we use \SI{20}{\ms} windows, an overlap of \SI{50}{\percent}, and a look-ahead of two frames resulting in an overall algorithmic delay of \SI{40}{\ms}.
We take \num{32} ERB bands, $f_\text{DF}=\SI{5}{\kHz}$, a DF order of $N=5$, and a look-ahead $l=2$ frames.
The loss parameters $\lambda_\text{spec}=1e3$ and $\lambda_\text{MR}=5e2$ are chosen so that both losses result in the same order of magnitude. The source code and a pretrained DeepFilterNet2 can be obtained at \url{https://github.com/Rikorose/DeepFilterNet}.

\subsection{Results}
\label{ssec:voicebank}
\begin{table*}[tbh]
  \caption{Objective results on Voicebank+Demand test set. Real-time factors (RTFs) are measured on a notebook Core i5-8250U CPU by taking the average over 5 runs. Unreported values of related work are indicated as ``-''.}
  \vspace{.2em}
  \begin{small}
  \label{tab:voicebank_test}
  \robustify\bfseries
  \sisetup{
    table-number-alignment = center,
    table-figures-integer  = 1,
    table-figures-decimal  = 2,
    table-auto-round = true,
    detect-weight = true
  }
  \begin{minipage}[c]{\textwidth}
    \centering
    \begin{tabular}{
        rl S S S[table-figures-decimal=2] S S S S S[table-figures-decimal=3]
      }%
      \toprule%
      & Model & Params $[\text{M}]$ & MACS $[\text{G}]$ & RTF & \text{PESQ} & CSIG  & CBAK & COVL & STOI \\%
      \cmidrule{2-10}%
      & Noisy                                          & \nr   & \nr   & \nr  & 1.97 & 3.34 & 2.44 & 2.63 & 0.921 \\%
      & RNNoise \cite{valin2018rnnoise}\footnote{Metrics and RTF measured with source code and weights provided at \url{https://github.com/xiph/rnnoise/}}%
                                                       & \bfseries0.06  & \bfseries0.04  & 0.027\footnote{Note, that RNNoise runs single-threaded}%
                                                                              & 2.33 & 3.40 & 2.51 & 2.84 & 0.922 \\%
      & NSNet2 \cite{braun2021towards}
                                                       & 6.168 & 0.43  & \bfseries0.022& 2.47 & 3.23 & 2.99 & 2.90 & 0.903 \\%
      & PercepNet \cite{valin2020perceptually}         & 8.0   & 0.8   & \nr  & 2.73 & \nr  & \nr  & \nr  & \nr  \\%
      & DCCRN \cite{hu2020dccrn} \footnote{RTF measured with source code provided at \url{https://github.com/huyanxin/DeepComplexCRN}} \footnote{Composite and STOI metrics provided by the same authors in \cite{lv2022sdccrn}}%
                                                       & 3.7   & 14.36 & 2.19 & 2.54 & 3.74 & 3.13 & 2.75 & 0.938 \\%
      & DCCRN+ \cite{lv2021dccrnplus}                  & 3.3   & \nr   & \nr  & 2.84 & \nr  & \nr  & \nr  & \nr  \\%
      & S-DCCRN \cite{lv2022sdccrn}                    & 2.34  & \nr   & \nr  & 2.84 & 4.03 & 3.43 & 2.97 & 0.940 \\%
      & FullSubNet+ \cite{chen2022fullsubnet+} \footnote{Metrics and RTF measured with source code and weights provided at \url{https://github.com/hit-thusz-RookieCJ/FullSubNet-plus}}%
                                                       & 8.67  & 30.06 & 0.545& 2.88 & 3.86 & 3.42 & 3.57 & 0.940 \\%
      & GaGNet \cite{li2022glance}\footnote{RTF measured with source code provided at \url{https://github.com/Andong-Li-speech/GaGNet/}}%
                                                       & 5.95  & 1.65  & 0.05 & 2.94 & \bfseries4.26 & 3.45 & 3.59 & \nr  \\%
      & DMF-Net \cite{yu2022dmf}                       & 7.84  & \nr   & \nr  & 2.97 & \bfseries4.26 & 3.52 & 3.62 & \bfseries0.944 \\%
      & FRCRN \cite{zhao2022frcrn}                     & 10.27 & 12.3  & \nr  & \bfseries3.21 & 4.23 & \bfseries3.64 & \bfseries3.73 & \nr \\%
      \cmidrule{2-10}
      & DeepFilterNet \cite{schroter2022deepfilternet} & \bfseries1.778 & \bfseries0.348 & 0.11 & 2.81 & 4.14 & 3.31 & 3.46 & 0.942 \\%
      \ldelim\{{5}{*}[\rot{proposed}] %
      & + Scheduling scheme                            & \bfseries1.778 & \bfseries0.348 & 0.11 & 2.92 & 4.22 & 3.39 & 3.58 & 0.941 \\%
      & \quad+ MR Spec-Loss                            & \bfseries1.778 & \bfseries0.348 & 0.11 & 2.98 & 4.20 & 3.41 & 3.60 & 0.942  \\%
      & \qquad+ Improved Data \& Augmentation          & \bfseries1.778 & \bfseries0.348 & 0.11 & 3.04 & \bfseries4.30 & 3.38 & 3.67 & 0.942 \\%
      & \quad\qquad+ Simplified DNN                    & 2.306 & 0.356 & \bfseries0.04 & \bfseries3.08 & \bfseries4.30 & \bfseries3.40 & \bfseries3.699& \bfseries0.9429 \\%
      & \qquad\qquad+ Post-Filter                      & 2.306 & 0.356 & \bfseries0.04 & 3.03 & 3.72 & 3.37 & 3.63& 0.941 \\%
      \cmidrule{2-10}%
    \end{tabular}
  \end{minipage}
  \end{small}
  \vspace{-.5em}
\end{table*}%
\begin{table}[tb]
  \vspace{-.7em}
  \caption{DNSMOS results on the DNS4 blind test set.}
  \vspace{.2em}
  \label{tab:dns4_blind}
  \robustify\bfseries
  \sisetup{
    table-number-alignment = center,
    table-figures-integer  = 1,
    table-figures-decimal  = 2,
    table-auto-round = true,
    detect-weight = true
  }
    \centering
    \begin{tabular}{
        l S S S
      }%
      \toprule%
      Model                                          & SIGMOS & BAKMOS & OVLMOS \\
      \midrule%
      Noisy                                          & 4.144  & 2.94   & 3.291 \\%
      RNNoise \cite{valin2018rnnoise}                & 3.884  & 3.694  & 3.378 \\%
      NSNet2 \cite{braun2021towards}                 & 3.866  & 4.210  & 3.585 \\%
      FullSubNet+ \cite{chen2022fullsubnet+}
                                                     & \bfseries4.215 & 4.117  & 3.751 \\%
      DeepFilterNet \cite{schroter2022deepfilternet} & 4.141  & 4.182  & 3.751 \\%
      DeepFilterNet2                                 & 4.196  & 4.427  & 3.882 \\%
      + Post-Filter                                  & 4.193  & \bfseries4.465  & \bfseries3.896 \\%
      \bottomrule%
    \end{tabular}
  \vspace{-.5em}
\end{table}%
We evaluate the speech enhancement performance of DeepFilterNet2 using the Valentini Voicebank+Demand test set \cite{valentini2016investigating}.
Therefore, we chose WB-PESQ \cite{ITU2007WBPESQ}, STOI \cite{taal2011stoi} and the composite metrics CSIG, CBAK, COVL \cite{hu2007composite}.
\Tab{tab:voicebank_test} shows DeepFilterNet2 results in comparison with other state-of-the-art (SOTA) methods.
One can find that DeepFilterNet2 achieves SOTA-level results while requiring a minimal amount of multiply-accumulate operation per second (MACS).
The number of parameters has slightly increased over DeepFilterNet (Sec. \ref{ssec:dnn}), but the network is able to run more than twice as fast and achieves a \num{0.27} higher PESQ score.
GaGNet \cite{li2022glance} achieves a similar RTF while having good SE performance.
However, it only runs fast when provided with the whole audio and requires large temporal buffers due to its usage of big temporal convolution kernels.
FRCRN \cite{zhao2022frcrn} is able to obtain best results in most metrics, but has a high computational complexity not feasible for embedded devices.

\Tab{tab:dns4_blind} shows DNSMOS P.835 \cite{reddy2022dnsmos} results on the DNS4 blind test set.
While DeepFilterNet \cite{schroter2022deepfilternet} was not able to enhance the speech quality mean opinion score (SIGMOS), with DeepFilterNet2 we obtain good results also for background and overall MOS values.
Moreover, DeepFilterNet2 comes relatively close to the minimum DNSMOS values that were used to select clean speech samples to train the DNS4 baseline NSNet2 (SIG=4.2, BAK=4.5, OVL=4.0) \cite{dubey2022dns4} further emphasizing its good SE performance.

\section{Conclusion}
\label{sec:conclusion}

In this work, we presented DeepFilterNet2, a low-complexity speech enhancement framework.
Taking advantage from DeepFilterNet's perceptual approach, we were able to further apply several optimizations resulting in SOTA SE performance.
Due to its lightweight architecture, it can be run on a Raspberry Pi 4 with a real-time factor of \num{0.42}.
In future work, we plan to extend the idea of speech enhancement to other enhancements, like correcting lowpass characteristics due to the current room environment.

\bibliographystyle{IEEEbib}
\bibliography{refs}

\begin{thebibliography}{10}

\bibitem{valin2020perceptually}
Jean-Marc Valin, Umut Isik, Neerad Phansalkar, Ritwik Giri, Karim Helwani, and
  Arvindh Krishnaswamy,
\newblock ``{A Perceptually-Motivated Approach for Low-Complexity, Real-Time
  Enhancement of Fullband Speech},''
\newblock in {\em INTERSPEECH 2020}, 2020.

\bibitem{schroter2022deepfilternet}
Hendrik Schröter, Alberto~N Escalante-B, Tobias Rosenkranz, and Andreas Maier,
\newblock ``{DeepFilterNet}: A low complexity speech enhancement framework for
  full-band audio based on deep filtering,''
\newblock in {\em IEEE International Conference on Acoustics, Speech and Signal
  Processing (ICASSP)}. IEEE, 2022.

\bibitem{zhao2022frcrn}
Shengkui Zhao, Bin Ma, Karn~N Watcharasupat, and Woon-Seng Gan,
\newblock ``{FRCRN}: Boosting feature representation using frequency recurrence
  for monaural speech enhancement,''
\newblock in {\em IEEE International Conference on Acoustics, Speech and Signal
  Processing (ICASSP)}. IEEE, 2022.

\bibitem{yu2022dmf}
Guochen Yu, Yuansheng Guan, Weixin Meng, Chengshi Zheng, and Hui Wang,
\newblock ``{DMF-Net}: A decoupling-style multi-band fusion model for real-time
  full-band speech enhancement,''
\newblock {\em arXiv preprint arXiv:2203.00472}, 2022.

\bibitem{li2022glance}
Andong Li, Chengshi Zheng, Lu~Zhang, and Xiaodong Li,
\newblock ``Glance and gaze: A collaborative learning framework for
  single-channel speech enhancement,''
\newblock {\em Applied Acoustics}, vol. 187, 2022.

\bibitem{schroeter2020clcnet}
Hendrik Schröter, Tobias Rosenkranz, Alberto Escalante~Banuelos, Marc
  Aubreville, and Andreas Maier,
\newblock ``{CLCNet}: {Deep} learning-based noise reduction for hearing aids
  using complex linear coding,''
\newblock in {\em IEEE International Conference on Acoustics, Speech and Signal
  Processing (ICASSP)}, 2020.

\bibitem{mack2019deep}
Wolfgang Mack and Emanu{\"e}l~AP Habets,
\newblock ``{Deep Filtering: Signal Extraction and Reconstruction Using Complex
  Time-Frequency Filters},''
\newblock {\em IEEE Signal Processing Letters}, vol. 27, 2020.

\bibitem{valentini2016investigating}
Cassia Valentini-Botinhao, Xin Wang, Shinji Takaki, and Junichi Yamagishi,
\newblock ``{Investigating RNN-based speech enhancement methods for
  noise-robust Text-to-Speech},''
\newblock in {\em SSW}, 2016.

\bibitem{dubey2022dns4}
Harishchandra Dubey, Vishak Gopal, Ross Cutler, Ashkan Aazami, Sergiy
  Matusevych, Sebastian Braun, Sefik~Emre Eskimez, Manthan Thakker, Takuya
  Yoshioka, Hannes Gamper, et~al.,
\newblock ``{ICASSP} 2022 deep noise suppression challenge,''
\newblock in {\em IEEE International Conference on Acoustics, Speech and Signal
  Processing (ICASSP)}. IEEE, 2022.

\bibitem{smith2017don}
Samuel~L Smith, Pieter-Jan Kindermans, Chris Ying, and Quoc~V Le,
\newblock ``Don't decay the learning rate, increase the batch size,''
\newblock {\em arXiv preprint arXiv:1711.00489}, 2017.

\bibitem{choi2021real}
Hyeong-Seok Choi, Sungjin Park, Jie~Hwan Lee, Hoon Heo, Dongsuk Jeon, and Kyogu
  Lee,
\newblock ``Real-time denoising and dereverberation wtih tiny recurrent
  u-net,''
\newblock in {\em International Conference on Acoustics, Speech and Signal
  Processing (ICASSP)}. IEEE, 2021.

\bibitem{reddy2021dns3}
Chandan~KA Reddy, Harishchandra Dubey, Kazuhito Koishida, Arun Nair, Vishak
  Gopal, Ross Cutler, Sebastian Braun, Hannes Gamper, Robert Aichner, and
  Sriram Srinivasan,
\newblock ``Interspeech 2021 deep noise suppression challenge,''
\newblock in {\em INTERSPEECH}, 2021.

\bibitem{valin2018rnnoise}
Jean-Marc Valin,
\newblock ``A hybrid dsp/deep learning approach to real-time full-band speech
  enhancement,''
\newblock in {\em 2018 IEEE 20th international workshop on multimedia signal
  processing (MMSP)}. IEEE, 2018.

\bibitem{braun2021towards}
Sebastian Braun, Hannes Gamper, Chandan~KA Reddy, and Ivan Tashev,
\newblock ``Towards efficient models for real-time deep noise suppression,''
\newblock in {\em IEEE International Conference on Acoustics, Speech and Signal
  Processing (ICASSP)}. IEEE, 2021.

\bibitem{hu2020dccrn}
Yanxin Hu, Yun Liu, Shubo Lv, Mengtao Xing, Shimin Zhang, Yihui Fu, Jian Wu,
  Bihong Zhang, and Lei Xie,
\newblock ``{DCCRN: Deep complex convolution recurrent network for phase-aware
  speech enhancement},''
\newblock in {\em INTERSPEECH}, 2020.

\bibitem{lv2022sdccrn}
Shubo Lv, Yihui Fu, Mengtao Xing, Jiayao Sun, Lei Xie, Jun Huang, Yannan Wang,
  and Tao Yu,
\newblock ``{S-DCCRN}: Super wide band dccrn with learnable complex feature for
  speech enhancement,''
\newblock in {\em IEEE International Conference on Acoustics, Speech and Signal
  Processing (ICASSP)}. IEEE, 2022.

\bibitem{lv2021dccrnplus}
Shubo Lv, Yanxin Hu, Shimin Zhang, and Lei Xie,
\newblock ``{DCCRN+: Channel-wise Subband DCCRN with SNR Estimation for Speech
  Enhancement},''
\newblock in {\em INTERSPEECH}, 2021.

\bibitem{chen2022fullsubnet+}
Jun Chen, Zilin Wang, Deyi Tuo, Zhiyong Wu, Shiyin Kang, and Helen Meng,
\newblock ``{FullSubNet+}: Channel attention fullsubnet with complex
  spectrograms for speech enhancement,''
\newblock in {\em IEEE International Conference on Acoustics, Speech and Signal
  Processing (ICASSP)}. IEEE, 2022.

\bibitem{ITU2007WBPESQ}
{ITU},
\newblock ``{Wideband extension to Recommendation P.862 for the assessment of
  wideband telephone networks and speech codecs},''
\newblock {\em ITU-T Recommendation P.862.2}, 2007.

\bibitem{taal2011stoi}
Cees~H Taal, Richard~C Hendriks, Richard Heusdens, and Jesper Jensen,
\newblock ``An algorithm for intelligibility prediction of time--frequency
  weighted noisy speech,''
\newblock {\em IEEE Transactions on Audio, Speech, and Language Processing},
  2011.

\bibitem{hu2007composite}
Yi~Hu and Philipos~C Loizou,
\newblock ``Evaluation of objective quality measures for speech enhancement,''
\newblock {\em IEEE Transactions on audio, speech, and language processing},
  2007.

\bibitem{reddy2022dnsmos}
Chandan~KA Reddy, Vishak Gopal, and Ross Cutler,
\newblock ``Dnsmos p. 835: A non-intrusive perceptual objective speech quality
  metric to evaluate noise suppressors,''
\newblock in {\em IEEE International Conference on Acoustics, Speech and Signal
  Processing (ICASSP)}. IEEE, 2022.

\end{thebibliography}

\end{document}